**Fluctuation-based evaluation of thermal conductivity using deviational phonon Monte Carlo simulations**

Takuma Hori[1,*)], Yuzuho Oki[1]

[1]*Department of Mechanical Systems Engineering, Tokyo University of Agriculture and Technology, Koganei, Tokyo, Japan*

*Author to whom correspondence should be addressed.

Electronic mail: hori@go.tuat.ac.jp.

Abstract

We develop a fluctuation-based Monte Carlo method for evaluating thermal conductivity within the phonon Boltzmann transport equation (BTE) framework. Starting from the energy-based variance-reduced formulation of the BTE, we derive a relation between the spatial spreading of deviational phonon energy and the macroscopic heat conduction equation. The resulting formulation expresses thermal conductivity in terms of the mean square displacement of deviational phonon particles, without imposing thermal reservoirs. Consequently, thermal conductivity along multiple directions can be evaluated from a single simulation. The method is demonstrated for crystalline silicon at room temperature in bulk, square nanowires, and anisotropic polycrystalline nanostructures. The calculated thermal conductivities agree with theoretical predictions, including the size dependence induced by phonon–boundary scattering and the mode-resolved contributions. For anisotropic polycrystalline structures, the method simultaneously resolves the directional thermal conductivities and captures their dependence on grain morphology. These results demonstrate the validity of the proposed fluctuation-based approach for evaluating thermal conductivity across different transport regimes and structural configurations.

1. Introduction

Thermal transport in crystalline semiconductors is governed primarily by phonons. In bulk crystals, phonon transport is described by intrinsic scattering processes, whereas in nanostructured materials, scattering from surfaces and interfaces become a dominant mechanism and substantially suppress the thermal conductivity. This thermal conductivity suppression is beneficial for applications requiring low thermal conductivity, such as thermoelectric materials, but is detrimental to heat dissipation in electronic devices. Therefore, accurate and efficient evaluation of thermal conductivity is essential for understanding and designing nanostructured materials in which phonon–boundary scattering plays a dominant role.

The phonon Boltzmann transport equation (BTE) provides a framework for describing phonon-mediated heat transport from the diffusive to the ballistic regime. Among the numerical approaches for solving the phonon BTE, the Monte Carlo (MC) method[1-3], which provides a stochastic solution, is suited to nanostructured systems because its particle-based formulation readily accommodates complex geometries and boundary conditions. Consequently, a wide variety of nanostructures including nanoporous[4-8] and polycrystalline nanostructures[9-13] have been investigated using the MC method and its various extensions. Among these approaches, the variance-reduced MC framework[14-16] has emerged as an efficient framework for solving the phonon BTE, substantially reducing statistical noise by explicitly simulating only deviations from equilibrium. In these MC methods, thermal conductivity is evaluated from the steady-state heat flux induced by thermal reservoirs imposed at the boundaries of the simulation domain, which establish a temperature gradient across the system.

Despite their broad applicability, such nonequilibrium MC approaches have several limitations arising from the explicit imposition of thermal reservoirs. First, because a temperature gradient is imposed along a prescribed direction, the thermal conductivity along different directions must be evaluated in separate simulations. This becomes costly for anisotropic materials and nanostructures, for which the thermal conductivity is also inherently anisotropic. Second, these methods are not readily applicable to periodic systems. When the phonon mean free path becomes comparable to or exceeds the simulation-domain size, the direct imposition of thermal reservoirs can introduce a strong dependence of the calculated thermal conductivity on the domain size. Specialized periodic boundary conditions have therefore been developed to reproduce the desired transport behavior, but determining an appropriate convergence criterion becomes nontrivial[15].

These limitations raise the question of whether thermal conductivity can be evaluated within the phonon BTE framework without explicitly imposing a temperature gradient. In molecular dynamics simulations, thermal conductivity can be obtained either from a nonequilibrium steady state generated by an imposed temperature gradient or from equilibrium fluctuations of the heat current using the Green–Kubo formalism[17,18]. The latter provides a natural treatment of periodic systems and does not require prescribed transport direction. However, corresponding fluctuation-based formulations remain underdeveloped within the phonon BTE framework. Existing approaches for evaluating thermal conductivity from phonon fluctuations have been limited to computationally expensive solutions of the conventional BTE[19-21]. In these studies, thermal conductivity was evaluated from equilibrium fluctuations of the phonon heat flux, following an Green–Kubo approach in molecular dynamics. Another study reported a method for evaluating the phonon mean free path from the mean square displacement of simplified gray phonon particles[22]. Although this approach is derived by analogy with the particle diffusion and phonon-gas kinetics, it is restricted to the gray approximation and has not been extended to phonons with mode-dependent properties. These shortcomings motivate the need for a fluctuation-based formulation that retains the computational efficiency of the MC framework while accounting for mode-dependent phonon properties and avoiding explicitly imposed thermal reservoirs.

In this study, we develop a fluctuation-based Monte Carlo method for evaluating thermal conductivity within the phonon BTE framework. Starting from the variance-reduced formulation of the phonon BTE, we establish a relation between the microscopic dynamics of deviational phonon particles and the macroscopic heat conduction equation, following derivations analogous to fluctuation-based formulations of transport coefficients in molecular dynamics methods. The resulting formulation determines the thermal conductivity from the mean square displacement of deviational phonon particles, without imposing a temperature gradient. Consequently, thermal conductivity along multiple directions can be evaluated from a single simulation, while periodic boundary conditions can be employed directly. We demonstrate the method for crystalline silicon at room temperature in bulk, nanowire, and polycrystalline nanostructures, thereby assessing its ability to evaluate thermal conductivity across different transport regimes and structural configurations.

## 2. Methodology

### 2-1. Theory for fluctuation-based evaluation of thermal conductivity

Thermal conductivity is evaluated by relating the spatial fluctuations of a deviational phonon particle to the macroscopic heat conduction equation. The derivation follows an approach analogous to that used to relate molecular trajectories to the diffusion coefficient and thermal conductivity in molecular dynamics simulations. Phonon transport is governed by the BTE, which describes the probability distribution function $f$ of a phonon mode with wave vector $\mathbf{k}$ and branch $s$ at position $\mathbf{r}$ and time $t$. Under the energy-based variance-reduced (EBVR) formulation and the relaxation time approximation, the BTE for phonon transport is expressed as Eq. (1)[14,15],

$$\frac{\partial e^{\mathrm{d}}_{\mathbf{k}s}(\mathbf{r},t)}{\partial t}+\mathbf{v}_{\mathbf{k}s}\cdot\frac{\partial e^{\mathrm{d}}_{\mathbf{k}s}(\mathbf{r},t)}{\partial \mathbf{r}}=-\frac{e^{\mathrm{d}}_{\mathbf{k}s}(\mathbf{r},t)-\left[e^{0}_{\mathbf{k}s}(T(\mathbf{r},t))-e^{0}_{\mathbf{k}s}(T_{\mathrm{base}})\right]}{\tau_{\mathbf{k}s}}, \tag{1}$$

where $e^{\mathrm{d}} = e\text{-}e^{0}(T_{\mathrm{base}}) = \hbar\omega f - \hbar\omega f^{0}(T_{\mathrm{base}})$, $f^{0}(T_{\mathrm{base}}) = [\exp(\hbar\omega/k_{\mathrm{B}}T_{\mathrm{base}})] - 1]^{-1}$ with $T_{\mathrm{base}}$ denoting base temperature. For simplicity, the subscripts $\mathbf{k}$ and $s$ are omitted for $\omega$, $e$, and $f$. Here, $\omega$, $\mathbf{v}$, and $\tau$ are phonon frequency, group velocity, and relaxation time, respectively, while $\hbar$ and $k_{\mathrm{B}}$ denote the reduced Planck constant and Boltzmann constant. When the temperature deviation $T(\mathbf{r}, t)$ - $T_{\mathrm{base}}$ is sufficiently small, the energy deviation on the right-hand side of Eq. (1) can be linearized as

$$e^{0}_{\mathbf{k}s}(T(\mathbf{r},t))-e^{0}_{\mathbf{k}s}(T_{\mathrm{base}})\approx[(T(\mathbf{r},t)-T_{\mathrm{base}}]\frac{de^{0}_{\mathbf{k}s}}{dT}. \tag{2}$$

Within this framework, we consider the spatial fluctuations in the energy density and temperature generated by introducing a single deviational phonon particle. The particle is introduced at **r**=**r**' at time $t$=0 with a deviational energy $\varepsilon$, into a system initially at a uniform temperature $T_{\text{base}}$, and its subsequent evolution is governed by the BTE in Eq. (1). The energy $\varepsilon$ can be chosen arbitrarily and does not affect the resulting thermal conductivity. The local temperature rise induced by the deviational particle is assumed to be sufficiently small so that the linearization in Eq. (2) remains valid. Because the energy-based formulation conserves deviational energy during phonon scattering and mode conversion, the total deviational energy remains equal to $\varepsilon$, while its spatial distribution evolves with time. To relate the microscopic fluctuation to macroscopic heat transport, we consider the anisotropic heat conduction equation,

$$\frac{\partial E(\mathbf{r},t)}{\partial t} = \sum_{\alpha=x,y,z} \kappa_\alpha \frac{d^2 T(\mathbf{r},t)}{dr_\alpha^2}, \tag{3}$$

where $E$, $T$, and $\kappa_\alpha$ denotes the energy density, temperature, and thermal conductivity in direction $\alpha$, respectively. The deviations of energy density $\tilde{E}$ and temperature $\tilde{T}$ from the base state are defined as

$$\tilde{E}(\mathbf{r},t) = E(\mathbf{r},t) - E_{\text{base}}, \tag{4}$$

$$\tilde{T}(\mathbf{r},t) = T(\mathbf{r},t) - T_{\text{base}}, \tag{5}$$

where $E_{\text{base}}$ is the energy density at temperature $T_{\text{base}}$. Substituting Eqs. (4) and (5) into

Eq. (3) yields

$$\frac{\partial \tilde{E}(\mathbf{r},\mathrm{t})}{\partial t} = \sum_{\alpha=x,y,z} \kappa_\alpha \frac{d^2 \tilde{T}(\mathbf{r},t)}{d r_\alpha^2}. \tag{6}$$

The initial condition corresponding to the introduction of a single deviational particle can be expressed with employing Dirac delta function $\delta$ as

$$\tilde{E}(\mathbf{r},0) = \varepsilon \delta(\mathbf{r}-\mathbf{r}'). \tag{7}$$

To solve Eq. (6) subject to the initial condition given by Eq. (7), these equations are reformulated in terms of phonon energy. First, the deviational energy density $\tilde{E}$ is expressed using phonon energies as, .

$$\tilde{E}(\mathbf{r},t) = \frac{1}{V_\mathrm{c}} \sum_{\mathbf{k},s} \hbar\omega [f(T(\mathbf{r},t)) - f(T_\mathrm{base})] = \frac{1}{V_\mathrm{c}} \sum_{\mathbf{k},s} [e(T(\mathbf{r},t)) - e(T_\mathrm{base})]. \tag{8}$$

Here, the summation is taken over all phonon modes (wave vectors and branches) within a crystal of volume $V_\mathrm{c}$. Furthermore, for sufficiently small temperature deviations, namely $T(\mathbf{r},t)$- $T_\mathrm{base} \approx 0$, the linearization in Eq. (2) is applicable. Consequently, Eq. (8) is approximated as

$$\tilde{E}(\mathbf{r},t) \approx \frac{1}{V_\mathrm{c}} \sum_{\mathbf{k},s} \frac{d e_{\mathbf{k}s}^0}{dT} [T(\mathbf{r},t) - T_\mathrm{base}] = \frac{1}{V_\mathrm{c}} \sum_{\mathbf{k},s} \frac{d e_{\mathbf{k}s}^0}{dT} \tilde{T}(\mathbf{r},t) \equiv c\tilde{T}(\mathbf{r},t), \tag{9}$$

where $c$ denotes the specific heat obtained by summing the mode-resolved ones over all phonon modes. By substituting Eq. (9) into Eq. (6), the heat conduction equation is rewritten in terms of the deviational energy density,

$$\frac{\partial \tilde{E}(\mathbf{r},\mathrm{t})}{\partial t}=\frac{1}{c}\sum_{\alpha=x,y,z}\kappa_\alpha\frac{d^2\tilde{E}(\mathbf{r},\mathrm{t})}{dr_\alpha^2}. \tag{10}$$

The differential equation of Eq. (10) can be solved using spatial Fourier transform defined as

$$F(\mathbf{q},t)=\frac{1}{(2\pi)^{3/2}}\int_{-\infty}^{\infty}\tilde{E}(\mathbf{r},t)e^{-i\mathbf{q}\cdot\mathbf{r}}d\mathbf{r}, \tag{11}$$

where **q** is the wave vector associated with the spatial Fourier transform. Applying this transformation to the governing equation in Eq. (10) and the initial condition in Eq. (7) yields

$$\frac{\partial F(\mathbf{q},t)}{\partial t}=-\frac{1}{c}\sum_{\alpha=x,y,z}\kappa_\alpha q_\alpha^2F(\mathbf{q},t), \tag{12}$$

$$F(\mathbf{q},0)=\frac{\varepsilon}{(2\pi)^{3/2}}e^{-i\mathbf{q}\cdot\mathbf{r}'}. \tag{13}$$

The solution of Eq. (12) satisfying the initial condition in Eq. (13), and its inverse spatial Fourier transform are given by

$$F(\mathbf{q},t)=\frac{\varepsilon}{(2\pi)^{\frac{3}{2}}}\exp\left(-i\mathbf{q}\cdot\mathbf{r}'-\frac{1}{c}\sum_{\alpha=x,y,z}\kappa_\alpha q_\alpha^2\,t\right), \tag{14}$$

$$\tilde{E}(\mathbf{r},t)=\varepsilon\prod_{\alpha=x,y,z}\left(\frac{c}{4\pi\kappa_\alpha t}\right)^{\frac{1}{2}}\exp\left[-\frac{c}{4\kappa_\alpha t}(r_\alpha-r_\alpha')^2\right]. \tag{15}$$

Equation (15) describes the spatial spreading of deviational energy. This spreading can be represented in terms of the motion of a deviational particle as

$$\tilde{E}(\mathbf{r},t)=\varepsilon P(\mathbf{r},t), \tag{16}$$

where $P(\mathbf{r},\ t)$ represents the probability density of finding the deviational particle at position $\mathbf{r}$ at time $t$. Combining Eqs. (15) and (16) yields the explicit form of the probability density function as

$$P(\mathbf{r},t) = \prod_{\alpha=x,y,z} \left(\frac{C}{4\pi\kappa_\alpha t}\right)^{\frac{1}{2}} \exp\left[-\frac{C}{4\kappa_\alpha t}(r_\alpha - r'_\alpha)^2\right]. \tag{17}$$

Defining the displacement from the initial position as $\boldsymbol{\xi} = \mathbf{r} - \mathbf{r}'$, the mean square displacement is given by

$$\langle \xi_\alpha^2 \rangle = \int_{-\infty}^{\infty} \xi_\alpha^2 P d\boldsymbol{\xi} = 2\int_0^{\infty} \xi_\alpha^2 \left(\frac{C}{4\pi\kappa_\alpha t}\right)^{\frac{1}{2}} \exp\left[-\frac{C}{4\kappa_\alpha t}\xi_\alpha^2\right] d\xi_\alpha$$
$$= \frac{2\kappa_\alpha t}{C}. \tag{18}$$

In the long-time limit, where the macroscopic transport becomes diffusive, the thermal conductivity is therefore obtained as

$$\kappa_\alpha = \lim_{t\to\infty} \frac{C}{2t} \langle \xi_\alpha^2 \rangle. \tag{19}$$

This formulation is analogous to the Einstein relation for diffusion, in which the transport coefficient is obtained from the long-time growth of the mean square displacement. Equation (19) constitutes the central relation of the present fluctuation-based formulation. The thermal conductivity in each direction is determined from the long-time growth rate of the mean square displacement of deviational particles multiplied by the total phonon specific heat. The directional average is consequently given by

$$\kappa = \lim_{t\to\infty} \frac{c}{6t}\langle\xi^2\rangle. \tag{20}$$

Equations (19) and (20) provide a practical means of evaluating thermal conductivity from the trajectories of deviational particles obtained in MC simulations, whose details are described in the next section. The associated statistical uncertainty is estimated from the amount of trajectory data recorded during the simulation. Suppose that particle coordinates are recorded at intervals of $\Delta t_o = n_o \Delta t_s$, where $\Delta t_s$ is the simulation time step and $n_o$ is the number of simulation steps between successive outputs. For a simulation comprising $N_s$ steps, the number of stored coordinate sets is $N_o = N_s/n_o$. Including the initial position, displacements are evaluated from $N_o+1$ coordinate data points. The displacement $\xi_\alpha(\Delta t_n)$ over a time interval $\Delta t_n = n\Delta t_o$ ($n = 1, \ldots, N_o$) can then be computed from $K = N_o - n + 1$ independent segments. For $n < K$, the standard deviation $\sigma_\alpha(\Delta t_n)$ of the mean square displacement is given by the expression derived in a previous study[23]. After removing the contribution associated with observation errors, which is justified by the absence of experimental measurement noise in numerical simulations, the resulting expression becomes

$$(\sigma_\alpha/\langle\xi_\alpha^2\rangle)^2 = \frac{1}{3nK^2}(4n^2K + 2K - n^3 + n). \tag{21}$$

For sufficiently large $n$ and $K$, which is satisfied for transport processes over long time scales, Eq. (21) reduces to

$$\sigma_\alpha/\langle\xi_\alpha^2\rangle = \left[\frac{n(4K-n)}{3K^2}\right]^{\frac{1}{2}}. \tag{22}$$

When the results from $N_\mathrm{p}$ statistically independent deviational particles are averaged, the statistical uncertainty decreases in proportion to $N_\mathrm{p}^{-1/2}$, giving

$$\sigma_\alpha/\langle\xi_\alpha^2\rangle = \left[\frac{n(4K-n)}{3K^2N_\mathrm{p}}\right]^{\frac{1}{2}}. \tag{23}$$

For the directional average displacement $\xi$, the corresponding standard deviation $\sigma$ is

$$\sigma/\langle\xi^2\rangle = \frac{1}{3}\left[\frac{n(4K-n)}{K^2N_\mathrm{p}}\right]^{\frac{1}{2}}. \tag{24}$$

## 2-2. Method for fluctuation-based Monte Carlo simulation

The displacements of phonons are obtained using a particle-based MC method to solve the BTE. Most aspects of the procedures follow the EBVRMC method developed in previous studies[14,15], with modifications to the recorded quantities and the boundary conditions. Specifically, whereas the previous approaches evaluate thermal conductivity from the heat flux, the present method records the coordinates of deviational phonon particles to obtain their mean square displacement. Furthermore, unlike conventional approaches that impose thermal reservoirs at the system boundaries, the present method does not require them and can therefore be directly applied with periodic boundary conditions.

The governing equation and phonon transport processes are identical to those employed in the EBVRMC framework. As the initial condition, a single deviational phonon particle is randomly placed within the simulation domain, whereas the previous approaches generate deviational particles through injection from thermal reservoirs. The phonon mode of the initial deviational particle is selected according to the mode-weighting procedure described below. After initialization, the phonon particle undergoes successive migration and intrinsic scattering events governed by the group velocity and relaxation time associated with its phonon mode. In addition to intrinsic scattering, phonon–boundary scattering is considered when structural interfaces are present. While intrinsic scattering randomly reassigns the phonon mode according to the mode weight, phonon–boundary scattering is assumed to be elastic and therefore changes only the propagation direction while preserving the phonon mode. In contrast, in a previous study[24], the frequency range up to the maximum phonon frequency was divided into 128 uniformly spaced bins, and all modes within each bin were treated as having identical energies. The present study does not employ this frequency-binning approximation. The simulations are continued for sufficiently long time to obtain statistically converged phonon trajectories.

When a phonon encounters an interface, it is either transmitted or reflected according to the interfacial transmission probability. At free surfaces, the transmission probability is zero, and all incident phonons are therefore reflected. At grain boundaries, the transmission probability depends on the structural disorder between the adjacent crystals and on the phonon mode, particularly its frequency. In the present study, the interfacial transmission probability $\tau_{int}$ is modeled using the following frequency-dependent function[25],

$$\tau_{\mathrm{int}} = \frac{1}{\gamma\omega/\omega_{\mathrm{max}} + 1}, \tag{25}$$

where $\omega_{max}$ denotes the maximum phonon frequency, and $\gamma$ is a dimensionless parameter characterizing the degree of interfacial disorder. A larger $\gamma$ corresponds to stronger interfacial disorder and hence lower transmission. We set $\gamma = 5$ for all grain boundaries, which is within the range considered realistic based on experimental observations[26] and atomistic Green's-function calculations [27]. For both transmission and reflection events, the propagation direction after the collision is assumed to be fully diffuse.

The probabilities of assigning phonon modes depend on the transport process. In all cases, phonon modes are weighted according to mode-specific heat $C_{\mathbf{k}s}$,

$$C_{\mathbf{k}s} \equiv \frac{1}{V_{\mathrm{c}}}\frac{de_{\mathbf{k}s}^{0}}{dT}. \tag{26}$$

For certain cases, the group velocity and relaxation time are also included in the weighting[15]. Specifically, the following mode-weighting functions are used: (a) $C_{\mathbf{k}s}$ for initialization with phonons randomly distributed in space, (b)$C_{\mathbf{k}s}/\tau_{\mathbf{k}s}$ for intrinsic scattering, and (c)$C_{\mathbf{k}s}\mathbf{v}_{\mathbf{k}s}\cdot\mathbf{n}H(\mathbf{v}_{\mathbf{k}s}\cdot\mathbf{n})$ for emission from thermal reservoirs. Here, **n** is the outward normal vector to the thermal reservoir surface, and $H$ is the Heaviside step function, which restricts the emitted phonons to those propagating away from the reservoir. For phonon–boundary diffuse scattering at structural interfaces, the weighting rule corresponds to case (c), with the additional constraint that scattering is elastic. These mode-weighting procedures assume that phonon properties are available throughout the first Brillouin zone.

In this study, silicon at room temperature is employed as the representative material. The phonon properties are taken from the example dataset provided in ALAMODE[28], for which the interatomic interaction are calculated using OpenMX[29]. The mesh in reciprocal space is 20 × 20 × 20, and the calculated bulk thermal conductivity is 145.1 W $m^{-1}K^{-1}$. After excluding phonon modes whose individual contributions to thermal conductivity are smaller than 0.001 W $m^{-1}$ $K^{-1}$, the calculated bulk thermal conductivity becomes 145.0 W $m^{-1}$ $K^{-1}$. This value is in reasonable agreement with experimentally reported thermal conductivity for silicon at room temperature[30]. The phonon dispersion relation is approximated as isotropic, and phonon propagation is therefore treated as spherically symmetric. Accordingly, the post-collision propagation direction at interfaces is determined according to the Lambert cosine law. Nevertheless, the present formulation can be extended to anisotropic materials by directly using phonon properties in reciprocal space, as demonstrated in previous studies[24,31,32].

In the fluctuation-based MC simulations, the timestep is set to $\Delta t_s$ = 0.02 ps, which is sufficiently smaller than the minimum phonon relaxation time of silicon at room temperature used in this study (1.1 ps). The total number of time steps is chosen to be $N_s$ = 100,000,000, corresponding to a total simulation time of 2,000 ns. This duration is sufficiently longer than the maximum phonon relaxation time of 1.7 ns. The output interval is set to $n_o$ = 1,000 time steps. For each set of structural and simulation conditions, $N_p$ = 10,000 independent deviational particles are simulated.

## 3. Results

### 3.1 Bulk thermal conductivity

We first evaluate the bulk thermal conductivity of silicon using the fluctuation-based MC method. Ten independent simulation ensembles are performed, with each ensemble consisting of $N_{\mathrm{p}}$ = 10,000 independently initialized deviational phonon particles. Figure 1 shows the mean square displacement normalized by the observation time interval, $\langle \xi^2 \rangle / \Delta t$, obtained from the MC simulations. According to Eq. (20), this quantity is directly proportional to the thermal conductivity. The shaded regions indicate the statistical uncertainty estimated from Eq. (24). The results show that $\langle \xi^2 \rangle / \Delta t$ approaches a constant value for observation time intervals longer than approximately 100 ns, suggesting that the diffusive regime assumed in Eq. (20) has been reached. This timescale is sufficiently longer than the maximum phonon relaxation time in silicon at room temperature, 1.7 ns, and thus the observed convergence is consistent with the transition from ballistic to diffusive transport. In contrast, the statistical uncertainty increases with increasing observation time interval because the number of available displacement segments decreases at longer observation intervals.

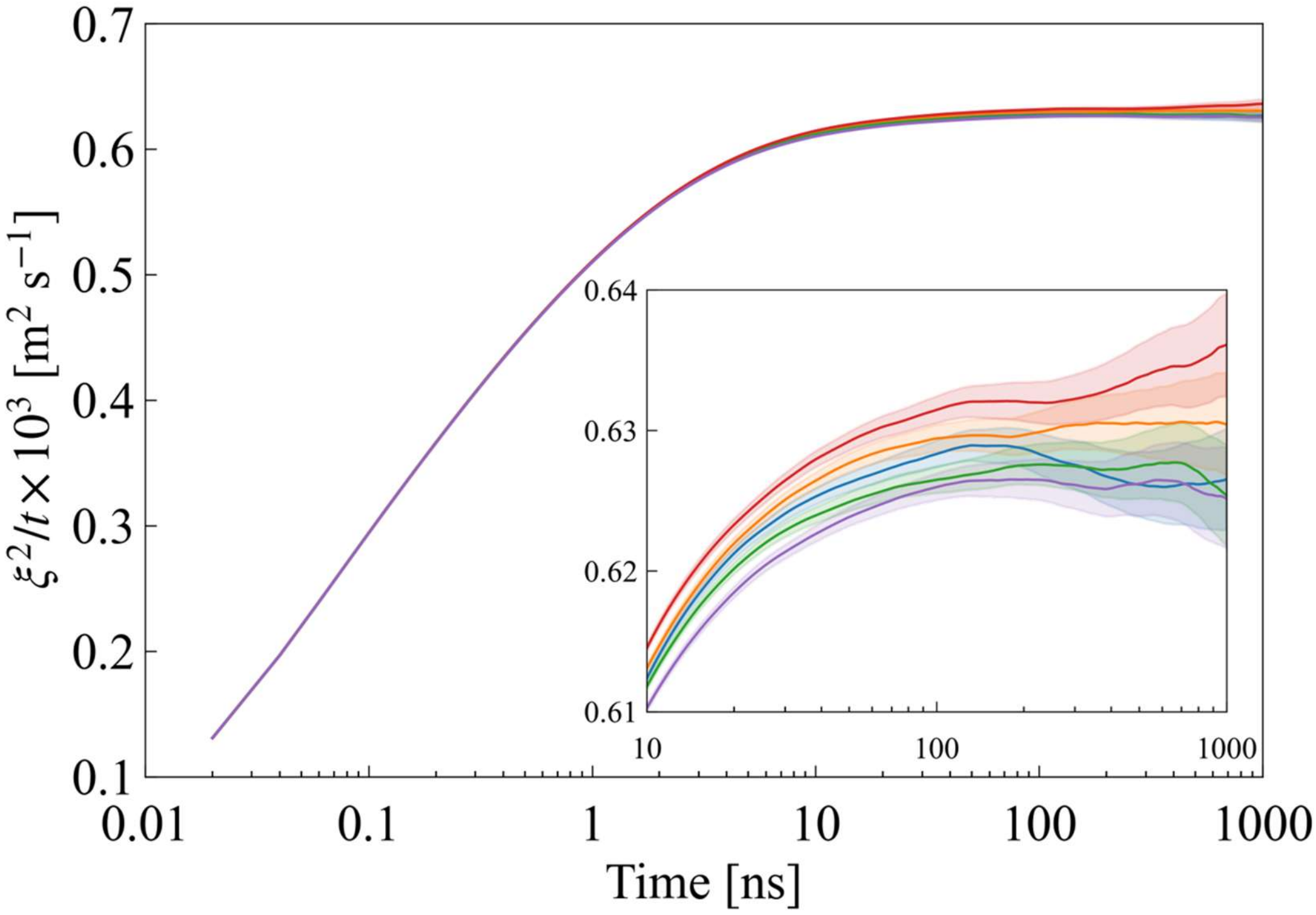

Fig. 1 Mean square displacement of deviational phonon particles as a function of observation time interval. Five of the ten independent simulation results are shown. The shaded regions indicate the standard deviations estimated from Eq. (24). The inset shows an enlarged view of the region at large observation time intervals.

The thermal conductivity is subsequently evaluated from the mean square displacement using Eq. (20). Figure 2 shows the thermal conductivity as a function of the observation time interval. For each interval, the mean and standard deviation are calculated from the ten independent simulation ensembles. Consistent with the behavior observed in Fig. 1, the calculated thermal conductivity converges with increasing observation time interval.

The standard deviation among the independent simulations is larger than the uncertainty shown in Fig. 1. This difference arises because Eq. (24) accounts only for the statistical uncertainty associated with displacements for a given set of particle trajectories, whereas the variation among independent ensembles additionally reflects the finite sampling of phonon modes. To assess the accuracy of the proposed method, the simulation results are compared with the theoretical bulk thermal conductivity given by

$$\kappa_{\mathrm{bulk}} = \sum_{\mathbf{k}s} \frac{1}{3} C_{\mathbf{k}s} v_{\mathbf{k}s}^{2} \tau_{\mathbf{k}s} \,, \tag{27}$$

As shown in Fig. 2, the calculated thermal conductivity converges to the theoretical value within the statistical uncertainty. This agreement demonstrates that the fluctuation-based MC method accurately reproduces the bulk thermal conductivity of silicon.

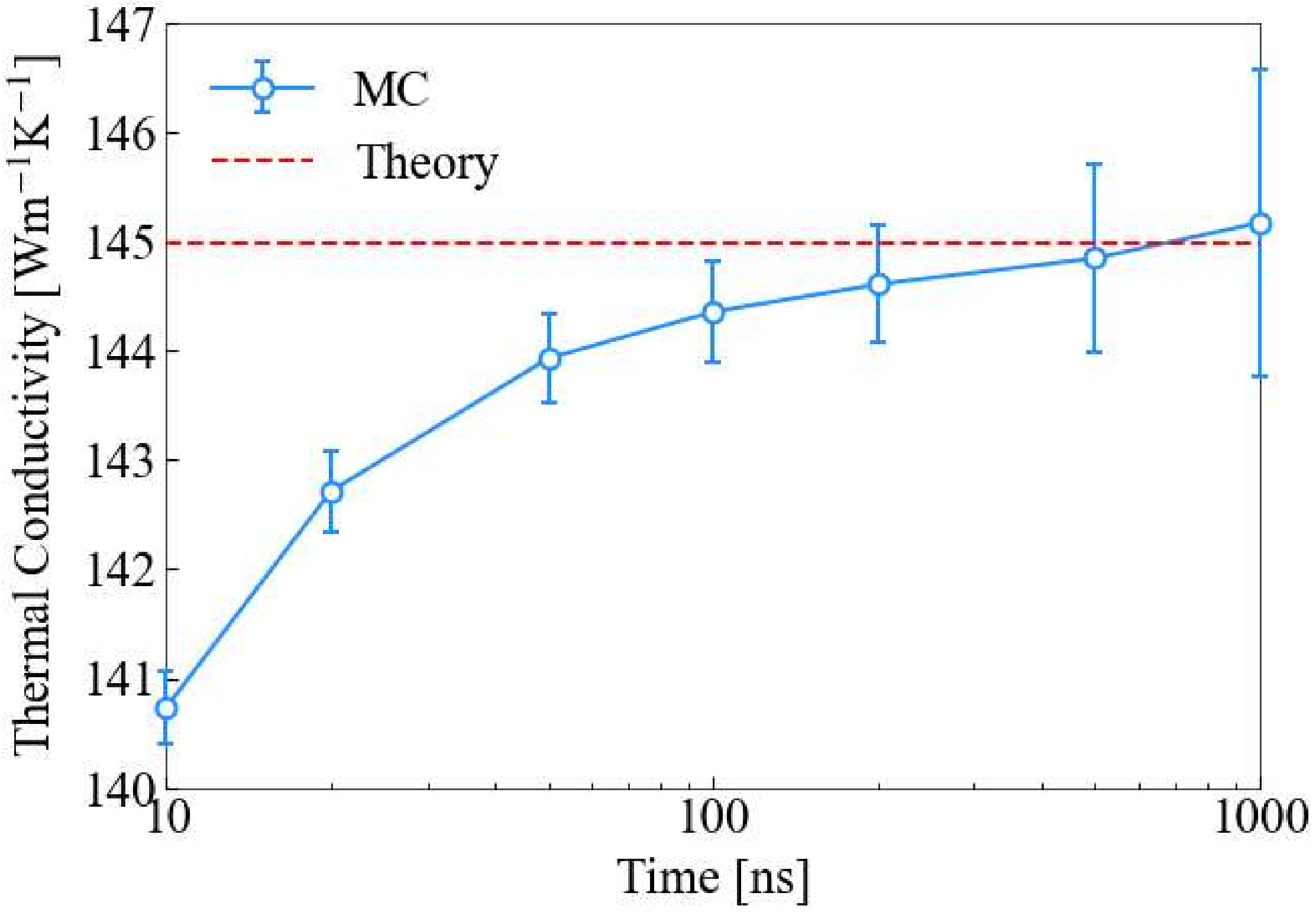

Fig. 2 Thermal conductivity as a function of observation time interval obtained from ten independent simulation ensembles. The error bars indicate the standard deviations among the ensembles. The dashed line represents the theoretical value given by Eq. (27).

We next evaluate the mode-resolved thermal conductivity using the fluctuation-based MC method. During a simulation, a deviational phonon particle undergoes multiple mode conversions through intrinsic scattering events. The total displacement of a phonon particle can therefore be decomposed into the displacements accumulated while the particle occupies individual phonon modes. Denoting the displacement accumulated while the phonon is in a given mode $\mathbf{k}s$ by $\xi_{\mathbf{k}s}$, the total squared displacement is

decomposed as

$$\xi^2 = \left(\sum_{\mathbf{k},s} \boldsymbol{\xi}_{\mathbf{k}s}\right)^2 = \sum_{\mathbf{k},s} \xi_{\mathbf{k}s}^2 + \sum_{\mathbf{k},s} \sum_{\mathbf{k}',s' \neq \mathbf{k},s} \boldsymbol{\xi}_{\mathbf{k}s} \cdot \boldsymbol{\xi}_{\mathbf{k}'s'} . \quad (28)$$

The first term on the right-hand side represents the self-correlation of displacements accumulated while the phonon particle occupied the same phonon mode, whereas the second term represents cross-correlations between displacements accumulated in different phonon modes. When taking the ensemble average, the cross-correlation terms vanish because displacements accumulated in different phonon modes are stochastically symmetry. Thus, substituting Eq. (28) into Eq. (20) yields the mode-resolved thermal conductivity $\kappa_{\mathbf{k}s}$ as

$$\kappa_{\mathbf{k}s} = \lim_{t \to \infty} \frac{C}{6t} \langle \xi_{\mathbf{k}s}^2 \rangle . \quad (29)$$

Figure 3 compares the mode-resolved thermal conductivity obtained from the fluctuation-based MC simulations with the corresponding theoretical values. The theoretical contribution of each mode is calculated directly from the integrand of Eq. (27), thus

$$\kappa_{\mathrm{bulk},\mathbf{k}s} = \frac{1}{3} C_{\mathbf{k}s} v_{\mathbf{k}s}^2 \tau_{\mathbf{k}s}. \quad (30)$$

The parity plot demonstrates excellent agreement between the MC results and the theoretical values over the entire range of mode contributions. This agreement confirms that the proposed method correctly reproduces not only the total bulk thermal conductivity but also the mode-resolved contribution to thermal conductivity.

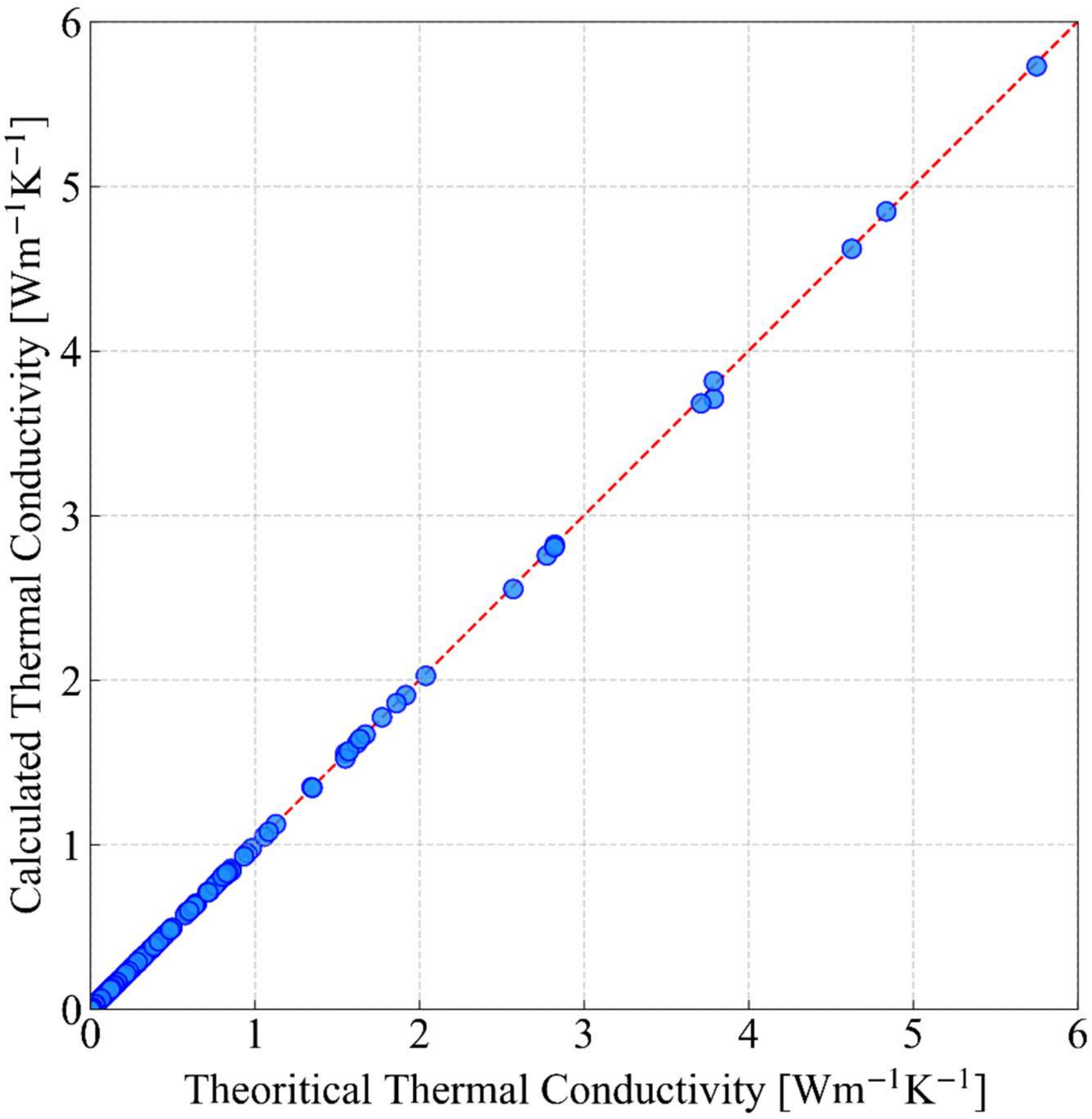

Fig. 3. Parity plot of the mode-resolved thermal conductivities calculated from the fluctuation-based MC simulations using Eq. (29) and the corresponding theoretical values from Eq. (30).

### 3.2 Thermal conductivity of nanowire

To demonstrate the applicability of the fluctuation-based MC method to nanostructured systems, we evaluate the thermal conductivity of square nanowires, where phonon–boundary scattering contributes to thermal resistance in addition to intrinsic scattering.

The side length of the square cross section $L_w$ is varied from 10 nm to 10,000 nm. Periodic boundary condition is imposed along the wire axis, thus the axial length of the simulation domain can be chosen independently of $L_w$. This configuration allows the thermal conductivity along the wire axis to be evaluated without introducing thermal reservoirs. Since only the thermal conductivity along the axial direction is considered for the nanowires, Eq. (19) is used for the evaluation.

The theoretical expression provides a reference for assessing whether the fluctuation-based MC method correctly captures the thermal conductivity reduction by the phonon – boundary scattering. In the ballistic transport limit, the boundary-limited mean free path $\Lambda_{bdy}$ of a square nanowire with side length $L_w$ is given by [33]

$$\Lambda_{\mathrm{bdy}} = 1.12 L_{\mathrm{w}}. \tag{31}$$

Assuming that intrinsic and boundary scattering processes act independently, the effective mean free path $\Lambda_{\mathrm{eff},\mathbf{k}s}$ can be approximated using Matthiessen's rule,

$$\frac{1}{\Lambda_{\mathbf{k}s,\mathrm{eff}}} = \frac{1}{\Lambda_{\mathbf{k}s}} + \frac{1}{\Lambda_{\mathrm{bdy}}}. \tag{32}$$

The corresponding theoretical thermal conductivity of the square nanowire is given by replacing $v_{\mathbf{k}s}\tau_{\mathbf{k}s}$ in Eq. (27) with $\Lambda_{\mathrm{eff},\mathbf{k}s}$.

Figure 4 shows the thermal conductivity of square nanowires as a function of the side length $L_w$. The mean and standard deviation are calculated from ten independent simulation ensembles, with the error bars smaller than the marker size. For small $L_w$, the boundary-limited mean free path becomes comparable to or smaller than the intrinsic mean free path of a large number of phonon modes, resulting in strong phonon–boundary scattering and reduced thermal conductivity. For large $L_w$, the boundary-limited mean free path exceeds the intrinsic mean free paths of the phonon modes, which reach approximately 7400 nm for the present phonon properties. Accordingly, the effective mean free paths approach their intrinsic values, and the resulting thermal conductivity approaches the bulk value obtained in the previous section. Thus, the size dependence shown in Fig. 4 represents a transition from ballistic to diffusive transport regimes. The MC results agree with the theoretical prediction based on Eq. (31) over the entire range of $L_w$ suggesting that the fluctuation-based MC method can capture the effect of phonon–boundary scattering across a wide range of transport regimes.

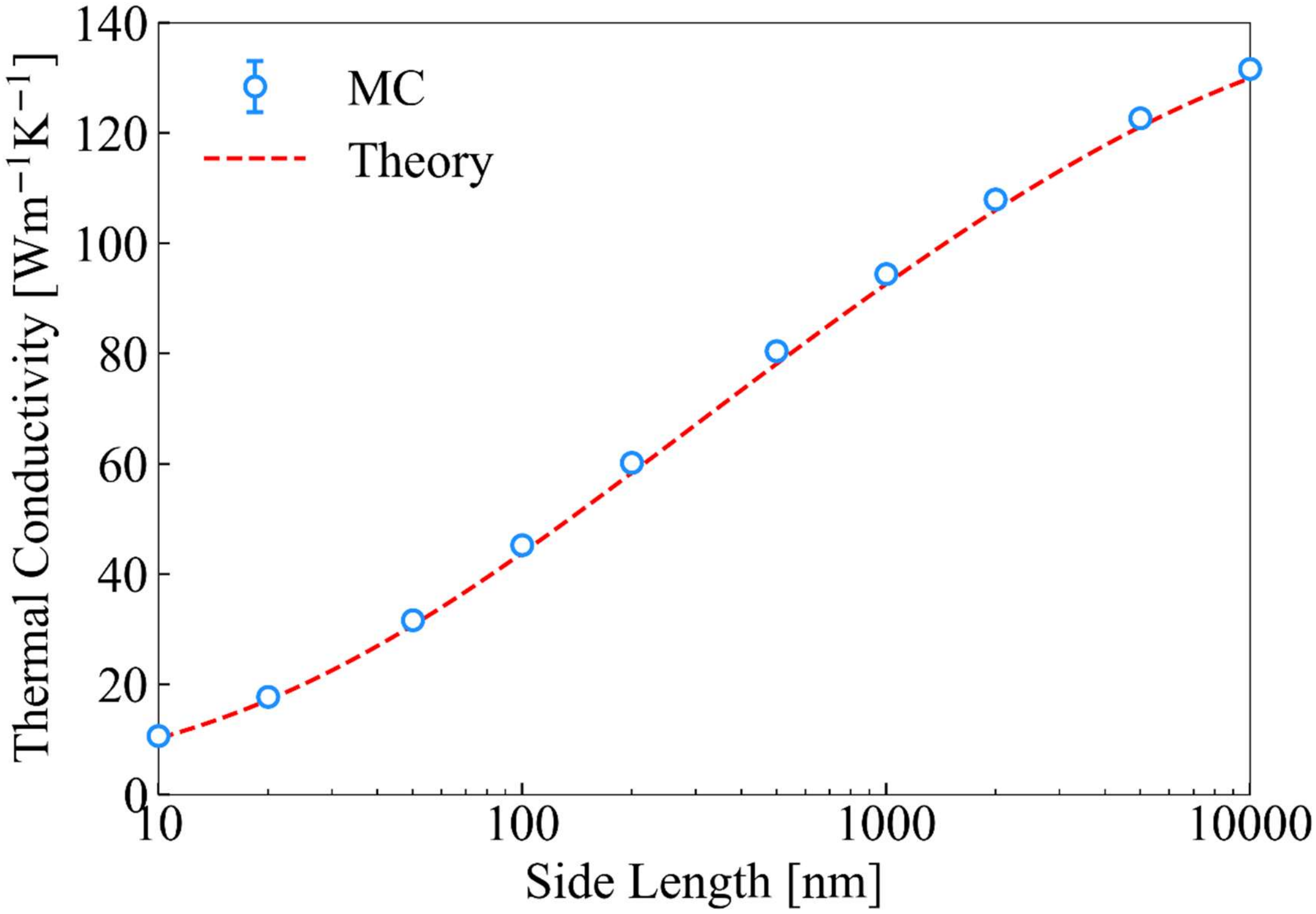

Fig. 4 Size-dependent thermal conductivity of square nanowires. The dashed line represents the theoretical prediction based on Matthiessen's rule, combining the boundary-limited and intrinsic phonon mean free paths, $\Lambda_{bdy}$ and $\Lambda_{\mathbf{k}s}$, respectively.

We next evaluate the mode-resolved thermal conductivity of the square nanowires to examine how phonon–boundary scattering modifies the contributions of individual phonon modes. Unlike Eq. (29), which gives the mode-resolved thermal conductivity averaged over the three spatial directions, the nanowire geometry requires the thermal conductivity along the wire axis to be considered. By applying the same mode decomposition to the $\alpha$ component of Eq. (28) and substituting it into Eq. (19), the mode-resolved thermal conductivity along the direction is obtained from the directional mean square displacement as

$$\kappa_{\mathbf{k}s,\alpha} = \lim_{t\to\infty} \frac{C}{2t} \langle \xi_{\mathbf{k}s,\alpha}^2 \rangle \, . \tag{33}$$

Figure 5 shows the mode-resolved thermal conductivity of square nanowires with side lengths of 10, 100, and 1000 nm. The acoustic and optical modes are shown separately. The mode-resolved thermal conductivity decreases as the nanowire size decreases, reflecting the increasing contribution of phonon–boundary scattering. The reduction is pronounced for the acoustic modes, as these modes possess larger intrinsic mean free path and are therefore more strongly affected when the boundary-limited mean free path becomes comparable to their intrinsic mean free paths. In contrast, the contributions of optical modes are less sensitive to the nanowire size because their shorter intrinsic mean free paths already limit their transport even in the bulk. The sum of the mode-resolved thermal conductivity reproduces the total thermal conductivity shown in Fig. 4, confirming the consistency between the mode-resolved and total thermal conductivity evaluations. These results demonstrate that the fluctuation-based MC method can resolve the modification of phonon transport at the mode level in nanostructures and thereby identify how phonons with different intrinsic transport properties contribute to the reduction in thermal conductivity caused by phonon–boundary scattering.

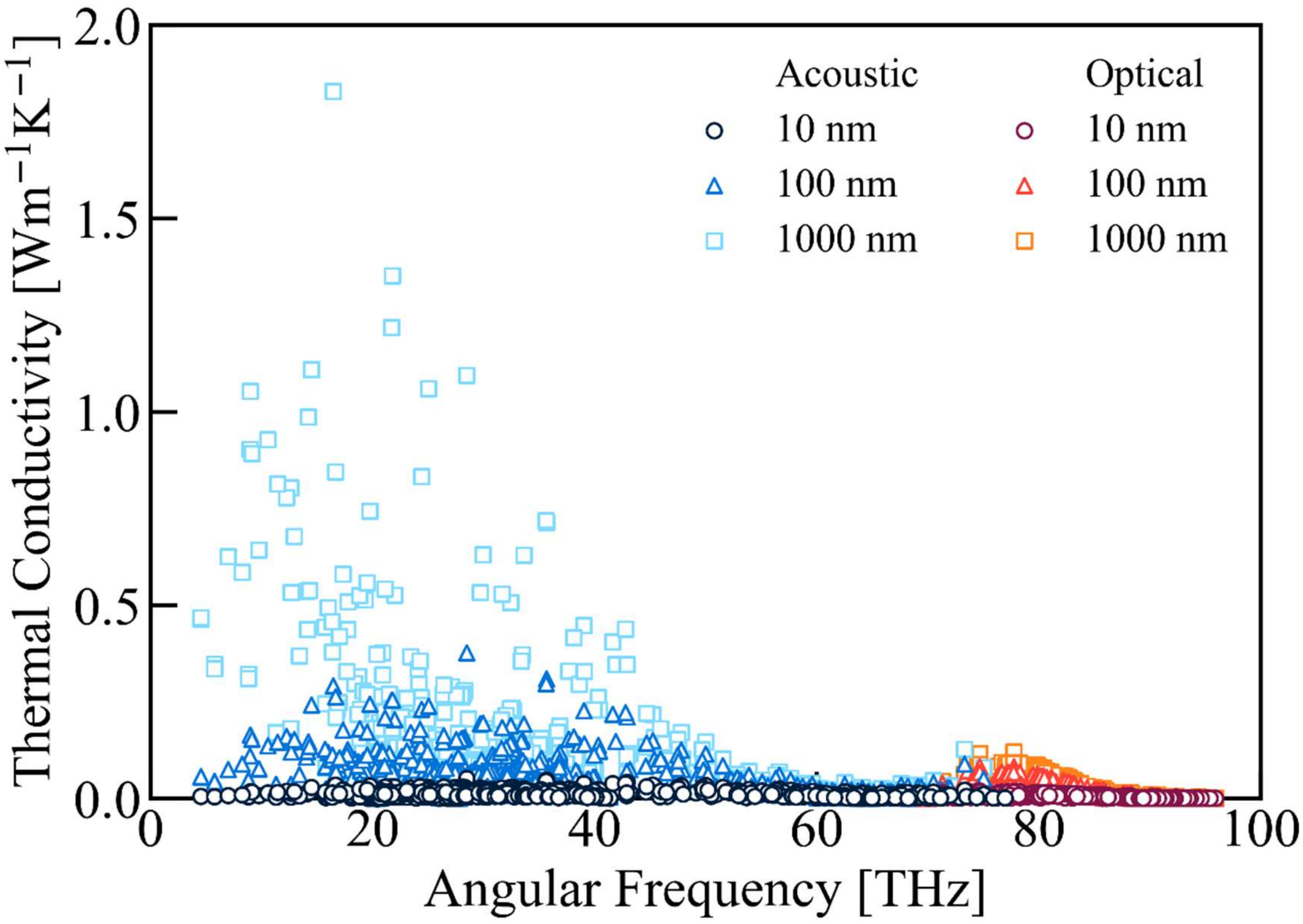

Fig. 5 Mode-dependent thermal conductivity of square nanowires with side lengths of 10, 100, and 1000 nm. The acoustic and optical modes are shown separately for each nanowire size.

3.3 Thermal conductivity of polycrystalline nanostructures

Using the developed fluctuation-based MC method, Eq. (19) enables the simultaneous evaluation of thermal conductivity along multiple spatial directions. To demonstrate this capability, we apply the proposed method to anisotropic polycrystalline nanostructures. The polycrystalline nanostructures are modeled using Voronoi tessellations, where the simulation domain is partitioned into regions associated with seed points. Voronoi tessellations have been widely employed in simulations of phonon transport in polycrystalline materials [9,10,13], as they can reproduce irregular grain morphologies without imposing overly simplified grain geometries such as cubic grains. To clearly demonstrate the capability of the proposed method to evaluate anisotropic thermal conductivity, we intentionally introduce anisotropy into the grain morphology. In Voronoi tessellation, randomly distributed seed points in a cubic domain generate statistically isotropic structures. Therefore, structural anisotropy is introduced by modifying the domain and seed-point coordinates along a specific direction, controlled by an anisotropy parameter $\eta$. First, $N_{\mathrm{poly}}$ seed points are randomly distributed within a simulation domain of dimensions $L_{\mathrm{poly}} \times L_{\mathrm{poly}} \times \eta L_{\mathrm{poly}}$. A three-dimensional Voronoi diagram is then generated under periodic boundary conditions. After the Voronoi diagram is constructed, the coordinates of both the simulation domain and the Voronoi vertices along the $z$ direction are uniformly scaled by a factor of $1/\eta$. The resulting structure is therefore contained in a cubic simulation domain of dimensions $L_{\mathrm{poly}} \times L_{\mathrm{poly}} \times L_{\mathrm{poly}}$, while the grain morphology becomes anisotropic. Specifically, the grains are elongated in the $z$ direction for $\eta < 1$ and flattened for $\eta > 1$.

The simulation domain size is fixed at $L_{\mathrm{poly}}$ = 2000 nm, and the number of grains is set to $N_{poly}$ = 8000. The corresponding characteristic grain size is therefore $D_{poly} = L_{poly}/N_{poly}^{1/3}$ = 100 nm. Representative polycrystalline nanostructures generated by this procedure are shown in Fig. 6. As intended, pronounced anisotropy in the grain morphology is observed except for $\eta$ = 1. The grain sizes appear to remain approximately unchanged for different values of $\eta$.

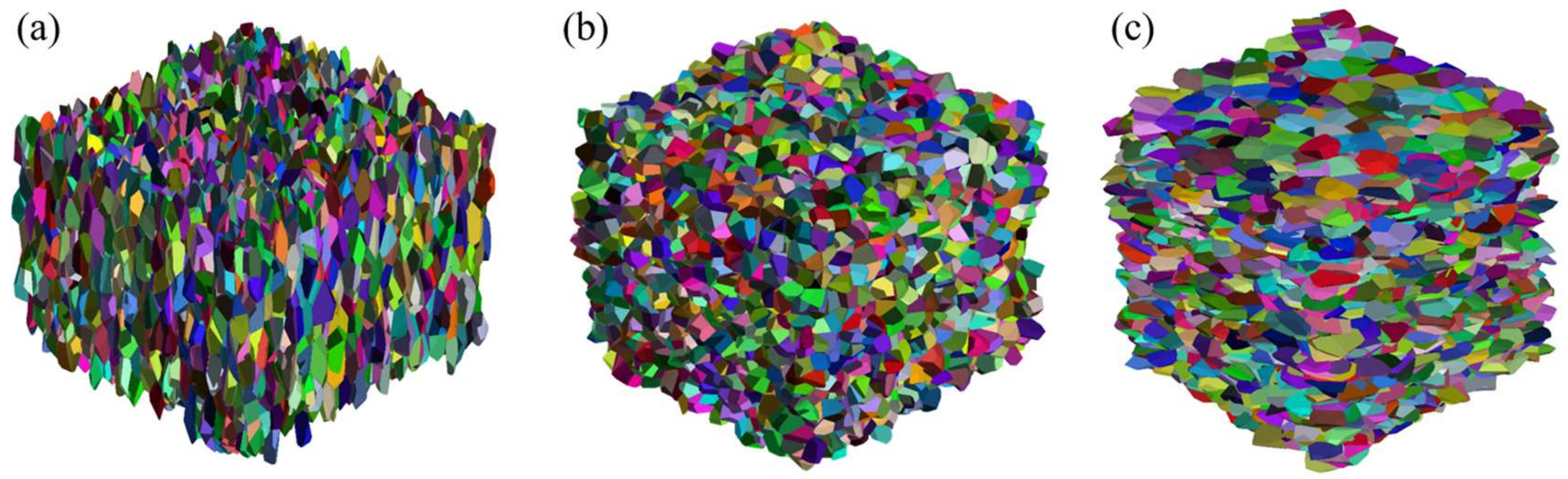

Fig. 6 Typical polycrystalline nanostructures modeled using Voronoi diagrams. The anisotropy parameter $\eta$ is set to (a) 0.25, (b) 1.0, and (c) 4.0. The number of grains is 8,000 in each case. The vertical direction corresponds to $z$ direction.

We next evaluate the thermal conductivity of the anisotropic polycrystalline nanostructures using fluctuation-based MC simulations. Ten independent simulations are performed for each value of anisotropic parameter $\eta$, and the mean values and standard deviations are calculated. Figure 7(a) shows the thermal conductivity along the three orthogonal directions, $\kappa_x$, $\kappa_y$, and $\kappa_z$ as a function of $\eta$, which is plotted on a logarithmic scale. The statistical uncertainties are smaller than the marker size for most data points. For $\eta = 1$, the thermal conductivities along the three directions agree with each other, consistent with the statistically isotropic structures shown in Fig. 6(b). As $\eta$ decreases below unity, $\kappa_z$ increases, whereas $\kappa_x$ and $\kappa_y$ decrease, as the grains become elongated along the $z$ direction, as shown in Fig. 6(a). The opposite tendency is observed when $\eta$ increases above unity, where the grains become flattened along the $z$ direction as shown in Fig. 6(c) resulting in a decrease in $\kappa_z$ and increases in $\kappa_x$ and $\kappa_y$.

Although the directional thermal conductivities exhibit approximately linear dependence on $\log\eta$, deviations from this trend are observed at the extreme values of $\eta$. For $\eta < 1$, the increase in $\kappa_z$ becomes progressively smaller as $\eta$ decreases, whereas the reduction in $\kappa_x$ and $\kappa_y$ exhibit an approximately linear dependence on $\log\eta$ over the investigated range. A corresponding behavior is also observed for $\eta > 1$. This behavior can be understood from the competing contributions of the grain dimensions in different directions. As phonon–grain-boundary scattering is assumed to be diffuse, the effective phonon mean free path is affected not only by the grain length along the transport direction but also by the spacing between grain boundaries transverse to the transport direction. For example, for $\eta < 1$, elongation of the grains along the $z$ direction increases the characteristic distance available for propagation in the $z$ direction and therefore tends to increase $\kappa_z$. At the same time, the corresponding reduction in the transverse grain dimensions increases the frequency of grain-boundary encounters for phonons propagating in the transverse directions. These competing geometrical contributions limit the increase in $\kappa_z$, resulting in the observed saturation-like behavior at small $\eta$.

To quantify the anisotropy in thermal conductivity, we define the thermal conductivity anisotropy as $2\kappa_z/(\kappa_x+\kappa_y)$. As shown in Fig. 7 (b), the thermal conductivity anisotropy exhibits an approximately linear dependence on $\log\eta$ over the investigated range. This result indicates that the anisotropic parameter provides a quantitative measure of the thermal conductivity anisotropy for the present polycrystalline nanostructures. Importantly, the three directional thermal conductivities are obtained from the same phonon-particle trajectories, without requiring additional simulations. This capability is particularly advantageous for anisotropic materials and nanostructures, for which conventional flux-based approaches generally require separate simulations for different transport directions. In addition, the simultaneous evaluation of directional thermal conductivities for thin-film structures with finite and periodically extended cross sections can be achieved by adopting a similar treatment to that reported in a previous study based on gray-phonon particles [34].

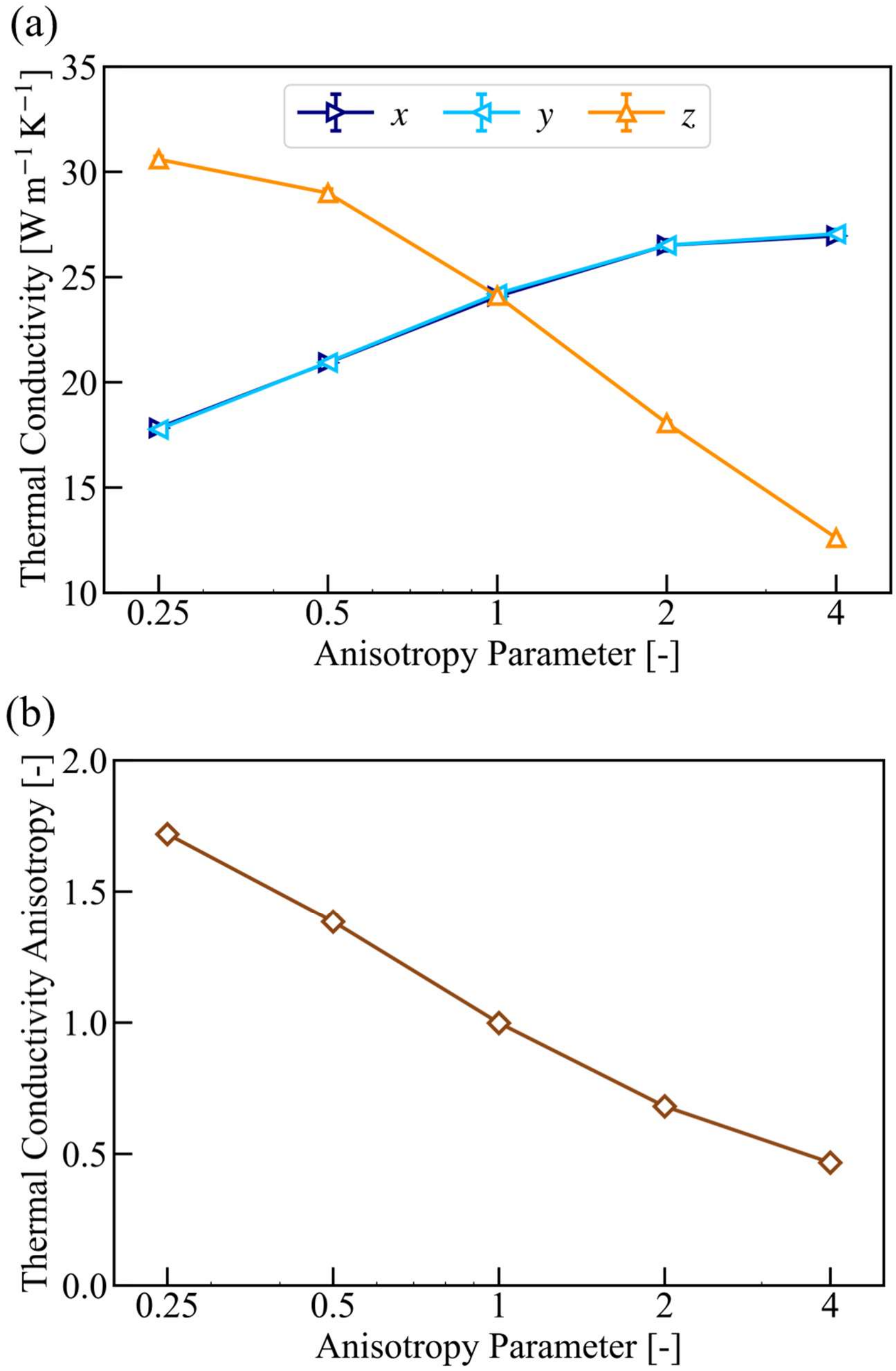

Fig. 7 (a) Thermal conductivity of anisotropic polycrystalline nanostructures along three orthogonal directions against anisotropy parameter $\eta$. (b) Corresponding thermal conductivity anisotropy, defined as $2\kappa_z/(\kappa_x+\kappa_y)$.

## 4. Conclusion

In this study, we develop a fluctuation-based Monte Carlo method for evaluating thermal conductivity within the phonon Boltzmann transport equation framework. By relating the spatial spreading of energies to the macroscopic heat conduction equation, we derive a relation that expresses thermal conductivity in terms of the mean square displacement of deviational phonon particles. Unlike conventional heat-flux-based approaches, the present method does not require thermal reservoirs or a prescribed transport direction. This feature allows multiple components of thermal conductivity to be evaluated from the same phonon-particle trajectories.

The proposed method is validated for crystalline silicon at room temperature in bulk and nanostructured systems. For bulk silicon, the calculated thermal conductivity converges to the theoretical value for sufficiently long observation times. In addition, the mode-resolved results reproduce the corresponding theoretical contributions. For square nanowires, the method captures the reduction in thermal conductivity caused by phonon–boundary scattering over a wide range of cross-sectional sizes and quantitatively agrees with the prediction based on Matthiessen's rule. For anisotropic polycrystalline nanostructures, the three directional components of thermal conductivity are simultaneously evaluated, and their anisotropy systematically reflects the anisotropy of the grain morphology. These results demonstrate that the proposed fluctuation-based Monte Carlo method provides a versatile approach for evaluating thermal conductivity, particularly in systems with anisotropic phonon transport.

## ACKNOWLEDGEMENTS

This study was financially supported by the JST FOREST Program Grant Number JPMJFR222 and the JSPS KAKENHI Grant Number JP26K00886.